\documentclass[aps,prd,preprintnumbers,superscriptaddress,amsmath,amssymb,nofootinbib,showpacs,twocolumn]{revtex4-1}%
\usepackage[dvips]{graphicx}
\usepackage{bm,latexsym,amsmath,amssymb,amsfonts,mathrsfs}
\usepackage{color}

\begin{document}

\title{Escape probability of the super-Penrose process}

\author{Kota Ogasawara}
\email[Email: ]{k.ogasawara@rikkyo.ac.jp}
\affiliation{Department of Physics, Rikkyo University, Toshima, Tokyo 171-8501, Japan}

\author{Tomohiro Harada}
\email[Email: ]{harada@rikkyo.ac.jp}
\affiliation{Department of Physics, Rikkyo University, Toshima, Tokyo 171-8501, Japan}

\author{Umpei Miyamoto}
\email[Email: ]{umpei@akita-pu.ac.jp}
\affiliation{RECCS, Akita Prefectural University, Akita 015-0055, Japan}

\author{Takahisa Igata}
\email[Email: ]{igata@rikkyo.ac.jp}
\affiliation{Department of Physics, Rikkyo University, Toshima, Tokyo 171-8501, Japan}

\begin{abstract}
We consider a head-on collision of two massive particles that move in
 the equatorial plane of an extremal Kerr black hole, which results in
 the production of two massless particles.
Focusing on a typical case, where both of the colliding particles have zero
 angular momenta, 
we show that a massless particle produced in such a collision can escape
 to infinity with arbitrarily large energy in the near-horizon limit of
 the collision point.
Furthermore, if we assume that the emission of the produced massless
 particles is isotropic 
in the center-of-mass frame but confined to the equatorial plane, 
the escape probability of
 the produced massless particle approaches $5/12$ and almost all escaping
 massless particles have arbitrarily large energy at infinity and 
an impact parameter approaching $2GM/c^2$, where $M$ is the mass of the black hole.
\end{abstract}

\preprint{RUP-16-26}
\maketitle

When two particles collide and produce two particles in the ergoregion
of a rotating black hole, one of the produced particles can escape to
infinity with energy larger than the total energy of the particles before the collision, which is called the collisional Penrose process \cite{PiranShahamKatz,PiranShaham}.
In 2009, Ba${\rm \tilde n}$ados, Silk, and West showed that the center-of-mass energy of two colliding particles can be arbitrarily large if they collide near the event horizon of an extremal Kerr black hole and one of the colliding particles has the critical value of the angular momentum \cite{BSW}.
This is called the Ba\~nados-Silk-West effect.
In recent years, this type of particle collision and the maximum of the energy-extraction efficiency have been investigated \cite{Bejger,HNM,Schnittman,Leiderschneider,OHM,HOM}.
If one of the colliding particles has the critical angular momentum and
both of the colliding particles come from infinity, the energy-extraction efficiency can reach $\simeq14$ \cite{Schnittman}, which is exactly given by $(2+\sqrt{3})^2$ \cite{Leiderschneider,OHM,HOM}.
Berti, Brito, and Cardoso showed that the efficiency of the collisional Penrose process becomes arbitrarily large, if two subcritical particles collide head on near the horizon \cite{BertiBritoCardoso}.
This is called the super-Penrose process.
In this case, a radially outward particle must be created near the horizon by some preceding process.
For example, particle emission and radiation from a collapsing star and an accretion disk as well as multiple scattering of infalling particles\footnote{Leiderschneider and Piran discuss that considering multiple collision of particles from infinity, the net energy extraction efficiency is 14 at most [E. Leiderschneider and T. Piran, 2015].}
can generate radially outward particles in the ergoregion.

The existence of near-extremal Kerr black holes is suggested by x-ray 
observations \cite{McClintock}. Therefore, 
the study of the super-Penrose process is important not only as 
one of the basic properties of an extremal Kerr black hole but
also from a point of view of observational astrophysics.

In this paper, we present an analytic formulation to investigate the energy-extraction efficiency and the escape probability of a typical super-Penrose process.
We show that the escape probability becomes $5/12$ and almost all the
escaping particles have arbitrarily large energy with an impact
parameter $b\simeq2M$ in the near-horizon limit.
We use the geometrized unit in which $c=G=1$.

The Kerr metric in the Boyer-Lindquist coordinates is given by
\begin{eqnarray}
ds^2&=&-\left(1-\frac{2Mr}{\rho^2}\right)dt^2-\frac{4Mar\sin^2\theta}{\rho^2}dtd\varphi+\frac{\rho^2}{\Delta}dr^2\nonumber\\
&&+\rho^2 d\theta^2+\left(r^2+a^2+\frac{2Ma^2r\sin^2\theta}{\rho^2}\right)\sin^2\theta d\varphi^2,\nonumber\\
\end{eqnarray}
where $\rho^2(r,\theta) := r^2+a^2\cos^2\theta$,
$\Delta(r):=r^2-2Mr+a^2$, and $M$ and $a \; (0 \leq a \leq M)$ are the mass and spin parameters, respectively.
The Kerr spacetime is stationary and axisymmetric with two Killing vectors $\partial_t$ and $\partial_\varphi$.
The conserved energy and angular momentum of a particle with a 4-momentum $p^\mu$ are given by
\begin{eqnarray}
E:=-g_{\mu\nu}(\partial_t)^\mu p^\nu~\mbox{and}~\;\;L:=g_{\mu\nu}(\partial_\varphi)^\mu p^\nu,
\end{eqnarray}
respectively.
The components of the 4-momentum are given by
\begin{eqnarray}
p^t&=&\frac{1}{\Delta}\left[\left(r^2+a^2+\frac{2Ma^2}{r}\right)E-\frac{2Ma}{r}L\right],\label{pt}\\
p^r&=&\sigma\sqrt{-2V(r)},\label{pr}\\
p^\theta&=&0,\label{ptheta}\\
p^\varphi&=&\frac{1}{\Delta}\left[\frac{2Ma}{r}E+\left(1-\frac{2M}{r}\right)L\right],\label{pphi}\\
V(r)&=&-\frac{Mm^2}{r}+\frac{L^2-a^2(E^2-m^2)}{2r^2}\nonumber\\
&&-\frac{M(L-aE)^2}{r^3}-\frac{E^2-m^2}{2},
\label{V}
\end{eqnarray}
where $\sigma={\rm sgn}(p^r)=\pm1$, $m$ denotes the mass of the particle, and the motion is assumed to be confined to the equatorial plane, $\theta=\pi/2$.
From now on, we concentrate on an extremal Kerr black hole, i.e., $a=M$.

A locally nonrotating frame (LNRF) is a tetrad basis associated with an observer who has zero angular momentum.
The transition that relates the components of a vector $V^\mu$ 
to the components in the LNRF $V^{(\alpha)}$ is given by
\begin{eqnarray}
V^{(\alpha)}=e^{(\alpha)}_\mu V^\mu,
\label{B.L._LNRF}
\end{eqnarray}
where $e^{(\alpha)}_\mu$ is the tetrad basis of the LNRF. 
According to Refs. \cite{Bardeen,PH}, the components of $e^{(\alpha)}_\mu$ are given by
\begin{eqnarray}
e^{(t)}_\mu&=&\left(\frac{r-M}{\sqrt{A(r)}},0,0,0\right),\label{et}\\
e^{(r)}_\mu&=&\left(0,\frac{r}{r-M},0,0\right),\label{er}\\
e^{(\theta)}_\mu&=&(0,0,r,0),\label{etheta}\\
e^{(\varphi)}_\mu&=&\left(-\frac{2M^2/r}{\sqrt{A(r)}},0,0,\sqrt{A(r)}\right),\label{ephi}
\end{eqnarray}
where $A(r):=r^2+M^2+2M^3/r$.
These satisfy $g_{\mu\nu}=\eta_{(\alpha)(\beta)}e^{(\alpha)}_\mu e^{(\beta)}_\nu$, where $\eta_{(\alpha)(\beta)}={\rm diag}(-1,1,1,1)$.
From Eqs. (\ref{pt})--(\ref{ephi}) the components of the 4-momentum in the LNRF are given by
\begin{eqnarray}
p^{(\alpha)}&=&\left(\frac{A(r)E-2M^2L/r}{(r-M)\sqrt{A(r)}},\frac{\sigma r\sqrt{-2V(r)}}{r-M},0,\frac{L}{\sqrt{A(r)}}\right).\nonumber\\
\label{p_LNRF}
\end{eqnarray}
We here consider the situation where two massive particles, named particles 1 and 2, collide head on and produce two massless particles, named particles 3 and 4.
For each particle $i$ ($i=1,2,3,4$), we write $p^{(\alpha)}$, $E$, $L$, $m$, and $\sigma$ as $p^{(\alpha)}_i$, $E_i$, $L_i$, $m_i$, and $\sigma_i$, respectively.
Without loss of generality, we assume that particle 1 moves radially outward, while particle 2 moves radially inward, i.e., $\sigma_1=-\sigma_2=1$.
We assume that particles 1 and 2 are marginally bound, of the same mass, and have zero angular momenta, i.e.,
\begin{eqnarray}
E_1=m_1=E_2=m_2=m,\;\;L_1=L_2=0.
\end{eqnarray}
In this case, $p^{(I)}_1+p^{(I)}_2$ ($I=r,\theta,\varphi$) vanish.
Thus, the LNRF coincides with the center-of-mass frame (CMF).
Therefore, $p^{(t)}_1+p^{(t)}_2$ yields the center-of-mass energy, that is,
\begin{eqnarray}
p^{(\alpha)}_1+p^{(\alpha)}_2=E_{\rm cm}(1,0,0,0),
\label{eq:p1+p2}
\end{eqnarray}
where
\begin{eqnarray}
E^2_{\rm cm}&:=&-\eta_{(\alpha)(\beta)}\left(p^{(\alpha)}_1+p ^{(\alpha)}_2\right)\left(p^{(\beta)}_1+p^{(\beta)}_2\right)\nonumber\\
&=&\frac{4m^2A(r_*)}{(r_*-M)^2}.
\label{E_cm}
\end{eqnarray}
We have denoted the radial position of the collision as $r_*$, where $M<r_*\leq2M$; i.e., we assume that the collision occurs in the ergoregion.
The local conservation of 4-momenta can be written as
\begin{eqnarray}
p^{(\alpha)}_1+p^{(\alpha)}_2=p^{(\alpha)}_3+p^{(\alpha)}_4.
\label{p_conserve}
\end{eqnarray}
From Eqs.~(\ref{eq:p1+p2}) and (\ref{p_conserve}), the 4-momenta of the two massless particles 3 and 4 can be written in the form
\begin{eqnarray}
p^{(\alpha)}_3&=&\frac{E_{\rm cm}}{2}(1,\cos\alpha,0,\sin\alpha),\label{p3_LNRF}\\
p^{(\alpha)}_4&=&\frac{E_{\rm cm}}{2}(1,-\cos\alpha,0,-\sin\alpha).\label{p4_LNRF}
\end{eqnarray}
This implies that the spatial velocity of particle 3  makes an angle $\alpha$ with $e^{(r)}_{\mu}$, where $-\pi\leq\alpha<\pi$.
From the components of the 4-momentum, we can obtain
\begin{eqnarray}
\sin\alpha&=&
\frac{p^{(\varphi)}_3}{\sqrt{\left(p^{(r)}_3\right)^2+\left(p^{(\varphi)}_3\right)^2}}=\frac{p^{(\varphi)}_3}{p^{(t)}_3}
=\frac{b(r_*-M)}{A(r_*)-2M^2b/r_*},\nonumber\\\label{sin}\\
\cos\alpha&=&
\frac{p^{(r)}_3}{\sqrt{\left(p^{(r)}_3\right)^2+\left(p^{(\varphi)}_3\right)^2}}=\frac{p^{(r)}_3}{p^{(t)}_3}\nonumber\\
&=&\frac{\sigma_3r_*\sqrt{A(r_*)}}{A(r_*)-2M^2b/r_*}\sqrt{1-\frac{b^2-M^2}{r^2_*}+\frac{2M(b-M)^2}{r^3_*}},\nonumber\\\label{cos}
\end{eqnarray}
where $b:=L_3/E_3$.

The components of the Killing vector $\partial_t$ in the LNRF are given by
\begin{eqnarray}
(\partial_t)^{(\alpha)}&=&\left(\frac{r-M}{\sqrt{A(r)}},0,0,-\frac{2M^2/r}{\sqrt{A(r)}}\right).
\label{Killing_t_LNRF}
\end{eqnarray}
From Eqs. (\ref{E_cm}), (\ref{p3_LNRF}), and (\ref{sin})--(\ref{Killing_t_LNRF}), $E_3$ is given by
\begin{eqnarray}
E_3=m\left(1+\frac{2M^2}{r_*(r_*-M)}\sin\alpha\right).
\label{E3def}
\end{eqnarray}
The energy-extraction efficiency in the collisional Penrose process is
defined as the ratio of the energy of the escaping particle to the total
energy of the colliding particles, i.e., 
\begin{eqnarray}
\eta:=\frac{E_3}{E_1+E_2}=\frac{1}{2}\left(1+\frac{2M^2}{r_*(r_*-M)}\sin\alpha\right).
\label{def_eta}
\end{eqnarray}
Because the Killing vector $\partial_t$ becomes spacelike inside the
ergoregion, the conserved energy $E_4$ can be negative.
In fact, if $\sin\alpha$ satisfies
\begin{eqnarray}
\sin\alpha>\frac{r_*(r_*-M)}{2M^2},
\end{eqnarray}
the energy-extraction efficiency can be larger than unity; i.e., the collisional Penrose process is at work.

\begin{center}
\begin{figure}[t]
\includegraphics[width=0.45\textwidth]{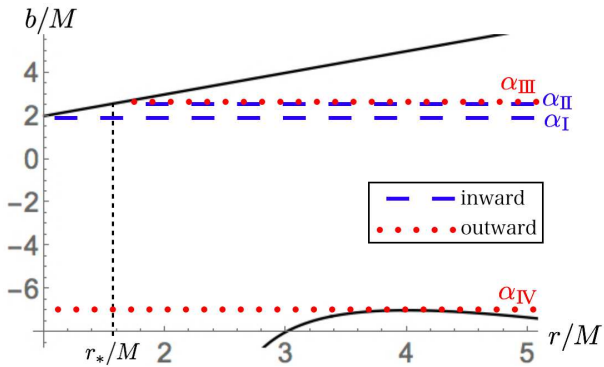}
\caption{The condition for particle 3 to escape to infinity is determined by investigating a radial turning point, i.e., $V=0$.
Solving $V=0$ for the impact parameter $b$, we obtain $b=b_\pm(r)$.
The radial turning points for particle 3 with $b=b_\pm(r)$ are plotted by the black solid lines.
The range of $b$ where particle 3 can escape from $r=r_*$ to infinity depends on whether it moves initially radially inward or outward.
If particle 3 moves initially inward, the maximum and minimum values of $b$ with which particle 3 can escape to infinity are given by the blue dashed lines labeled $\alpha_{\rm I\hspace{-.1em}I}$ and $\alpha_{\rm I}$.
The values of the blue dashed lines labeled $\alpha_{\rm I\hspace{-.1em}I}$ and $\alpha_{\rm I}$ are given by $b_+(r_*)/M$ and $2$, respectively.
If particle 3 moves initially outward, the maximum and minimum values of $b$ with which particle 3 can escape to infinity are given by the red dotted lines labeled $\alpha_{\rm I\hspace{-.1em}I\hspace{-.1em}I}$ and $\alpha_{\rm I\hspace{-.1em}V}$.
The values of the red dotted lines labeled $\alpha_{\rm I\hspace{-.1em}I\hspace{-.1em}I}$ and $\alpha_{\rm I\hspace{-.1em}V}$ are given by $b_+(r_*)/M$ and $-7$, respectively.
Please notice that the blue dashed line labeled $\alpha_{\rm I\hspace{-.1em}I}$ and the red dotted line labeled $\alpha_{\rm I\hspace{-.1em}I\hspace{-.1em}I}$ take the same value, i.e., $\alpha_{\rm I\hspace{-.1em}I}=\alpha_{\rm I\hspace{-.1em}I\hspace{-.1em}I}$.}
\label{PlotV}
\end{figure}
\end{center}

\begin{center}
\begin{figure}[t]
\includegraphics[width=0.5\textwidth]{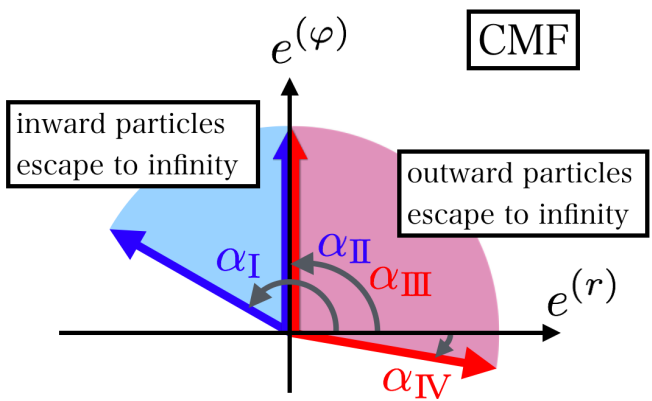}
\caption{The schematic diagram of the critical angles and the escape cone for particle 3, i.e., the range of angles in the CMF along which particle 3 must be emitted to escape from $r=r_*$ to infinity.
The colored arrows denote the spatial velocities of particle 3 with the critical angles.
The colored sectors denote the escape cone for particle 3.
The escape probability is given by the ratio of the angle of the escape cone to $2\pi$.
If particle 3 moves initially inward, the maximum and minimum values of $\alpha$ with which particle 3 can escape to infinity are given by $\alpha_{\rm I}$ and $\alpha_{\rm I\hspace{-.1em}I}$.
Particle 3 emitted radially inward in the range $\alpha\in(\alpha_{\rm I\hspace{-.1em}I},\alpha_{\rm I})$ bounces back at the turning point and can escape to infinity.
If particle 3 moves initially outward, the maximum and minimum values of $\alpha$ with which particle 3 can escape to infinity are given by $\alpha_{\rm I\hspace{-.1em}I\hspace{-.1em}I}$ and $\alpha_{\rm I\hspace{-.1em}V}$.
Particle 3 emitted radially outward in the range $\alpha\in(\alpha_{\rm I\hspace{-.1em}V},\alpha_{\rm I\hspace{-.1em}I\hspace{-.1em}I}]$ can directly escape to infinity.
Since $\alpha_{\rm I\hspace{-.1em}I}=\alpha_{\rm I\hspace{-.1em}I\hspace{-.1em}I}$, we represent the spatial velocities of particle 3 with $\alpha_{\rm I\hspace{-.1em}I}$ and $\alpha_{\rm I\hspace{-.1em}I\hspace{-.1em}I}$ as the one arrow
 colored half blue and half red.}
\label{R0L0}
\end{figure}
\end{center}

The condition for particle 3 to escape to infinity is determined by investigating a radial turning point, i.e., $V=0$.
Solving $V=0$ for the impact parameter $b$, we obtain $b=b_\pm$, where
\begin{eqnarray}
b_+(r)&:=&r+M,\label{b+}\\
b_-(r)&:=&-\left(r+M+\frac{4M^2}{r-2M}\right).\label{b-}
\end{eqnarray}
This means that particle 3 with impact parameter $b=b_\pm(r)$ has a turning point at $r$.
The numerical plot of $b=b_\pm(r)$ is given in Fig. \ref{PlotV}.
For $2M<b\leq b_+(r_*)$, particle 3 can escape to infinity irrespective of the sign of the initial radial velocity.
On the other hand, for $-7M<b\leq2M$, particle 3 can escape to infinity only if it moves initially outward.
We define critical angles as the boundaries of the impact parameter to
escape to infinity as follows:
\begin{eqnarray}
\alpha_{\rm I}
&:=&\alpha(\sigma_3=-1,\;r=r_*,\;b=2M),\label{alpha1}\\
\alpha_{\rm I\hspace{-.1em}I}
&:=&\alpha(\sigma_3=-1,\;r=r_*,\;b=b_+(r_*)),\label{alpha2}\\
\alpha_{\rm I\hspace{-.1em}I\hspace{-.1em}I}
&:=&\alpha(\sigma_3=1,\;r=r_*,\;b=b_+(r_*)),\label{alpha3}\\
\alpha_{\rm I\hspace{-.1em}V}
&:=&\alpha(\sigma_3=1,\;r=r_*,\;b=-7M).\label{alpha4}
\end{eqnarray}
From Eqs. (\ref{sin}), (\ref{b+}), (\ref{alpha2}), and (\ref{alpha3}), we can find
\begin{eqnarray}
\sin\alpha_{\rm I\hspace{-.1em}I}
=\sin\alpha_{\rm I\hspace{-.1em}I\hspace{-.1em}I}
=1.
\end{eqnarray}
From Eqs. (\ref{sin}), (\ref{cos}), (\ref{alpha1}), and (\ref{alpha4}), we can find
\begin{eqnarray}
\sin\alpha_{\rm I}>0,\;\cos\alpha_{\rm I}<0,\;\sin\alpha_{\rm I\hspace{-.1em}V}<0,\;\cos\alpha_{\rm I\hspace{-.1em}V}>0.
\end{eqnarray}
Therefore, we can find
\begin{eqnarray}
\alpha_{\rm I\hspace{-.1em}V}<\alpha_{\rm I\hspace{-.1em}I\hspace{-.1em}I}=\alpha_{\rm I\hspace{-.1em}I}<\alpha_{\rm I}.
\end{eqnarray}
This implies that the escape cone for particle 3, i.e., the range of angles in the CMF along which particle 3 must be emitted to escape to infinity, is given by
\begin{eqnarray}
\alpha\in(\alpha_{\rm I\hspace{-.1em}V},\alpha_{\rm I}).
\end{eqnarray}
The explicit form of the critical angles is given by
\begin{eqnarray}
\alpha_{\rm I}&=&\frac{5\pi}{6}+O(\epsilon),\label{sol_alpha1}\\
\alpha_{\rm I\hspace{-.1em}I}&=&\alpha_{\rm I\hspace{-.1em}I\hspace{-.1em}I}=\frac{\pi}{2},\label{sol_alpha23}\\
\alpha_{\rm I\hspace{-.1em}V}&=&-\frac{7}{18}\epsilon+O(\epsilon^2),\label{sol_alpha4}
\end{eqnarray}
where we have assumed that the radial position of near-horizon collision $r_*$ is given by
\begin{eqnarray}
r_*=\frac{M}{1-\epsilon},\;\;0<\epsilon\ll1.
\label{r_collision}
\end{eqnarray}
See Fig. \ref{R0L0} for the schematic diagram of the situation.

The escape probability is given by the ratio of the angle of the 
escape cone to $2\pi$ if we assume that the emission of the produced massless particles is isotropic in the CMF and confined to the equatorial plane.
It is given by
\begin{eqnarray}
P:=\frac{\alpha_{\rm I}-\alpha_{\rm I\hspace{-.1em}V}}{2\pi}.
\label{def_P}
\end{eqnarray}
From Eqs. (\ref{sol_alpha1}), (\ref{sol_alpha4}), and (\ref{def_P}), we can find
\begin{eqnarray}
P\to\frac{5}{12},
\end{eqnarray}
in the limit $\epsilon\to0$.

From Eqs. (\ref{def_eta}) and (\ref{r_collision}), if $\sin\alpha=O(1)$, $\eta$ is given by
\begin{eqnarray}
\eta=\frac{E_3}{2m}=\frac{\sin\alpha}{\epsilon}+O(1)
\end{eqnarray}
and this diverges to infinity in the limit $\epsilon\to0$.
Only if $\sin\alpha=O(\epsilon)$, $\eta$ turns out to be finite.
If particle 3 is emitted along the angle $\alpha_{\rm I\hspace{-.1em}V}$, $\eta$ becomes
\begin{eqnarray}
\eta=\frac{E_3}{2m}=\frac{1}{9}+O(\epsilon).
\end{eqnarray}
This means that the proportion of the particles with finite energy is minuscule.
In other words, almost all the particles which escape to infinity have arbitrarily large energy.

From Eqs.~(\ref{sin}) and (\ref{r_collision}), the relation among $\alpha$, $b$, and $\epsilon$ is given by
\begin{eqnarray}
2M-b=\frac{b}{2\sin\alpha}\epsilon+O(\epsilon^{2}),
\end{eqnarray}
if $\sin\alpha=O(1)$.
Therefore, almost all the particles which escape to infinity have an impact parameter $b=2M+O(\epsilon)$.
We assumed that two colliding particles are marginally
bound, of the same mass, and have zero angular momenta.
In this case, the spatial components of the total 4-momentum of the colliding particles vanish in the LNRF, and hence the LNRF coincides with the CMF.
We computed the escape cone, i.e., the range of angles in the CMF along which particles must be emitted to escape to infinity, and computed the escape probability.
We showed that the escape probability approaches $5/12$ and almost all the escaping particles have arbitrarily large energy with an impact parameter $b\to 2M$ in the near-horizon limit of the collision point.
We would suggest that very high-energy particles with impact parameter $b\simeq2M$ are produced by super-Penrose process and can be in principle observed from rapidly rotating black holes in astrophysics.
Once a super-Penrose process occurs, the rotation of black hole ceases to be extremal due to its backreaction.
However, the black hole can be spun up to near-extremal rotation by further mass accretion, and super-Penrose process can be at work once again.

In this paper, we focus on a special set of parameter values of the colliding particles.
However, we would like to stress that it is not a singular example but a typical one which provides a picture that we believe is representative of more general collisions in the super-Penrose process.
Even if we do not assume that two colliding particles are marginally bound, of the same mass, or have zero momenta, the escape probability takes a value of the order of unity, and almost all the escaping particles have arbitrarily large energy except for the sets of fine-tuned parameter values.
The work in this direction is in preparation.

The authors would like to thank K. Tanabe.
This work was supported by JSPS KAKENHI Grants No. JP 26400282 (T.H.) and No. JP 15K05086 (U.M.)


\end{document}